\documentstyle[twocolumn,aps]{revtex}
\newcommand{\half}{\frac{1}{2}}
\begin{document}
\draft
\title{
Reply to Comment by Holas and March}
\author{R. K. Nesbet}
\address{
IBM Almaden Research Center,
650 Harry Road,
San Jose, CA 95120-6099, USA}
\date{\today}
\maketitle
\begin{center}For {\em Phys.Rev.A} \end{center}
\begin{abstract}
The accompanying Comment by A. Holas and N. H. March 
[Phys. Rev. A {\bf 66}, 066501 (2002)] is concerned with
the issue of whether or not kinetic energy can be represented by an
effective local potential, as required for an exact Thomas-Fermi theory
equivalent to Kohn-Sham density-functional theory.  They dispute  
[R.K. Nesbet, Phys. Rev. A {\bf 65}, 010502(R) (2001)],
which concludes that for more than two electrons the use by Kohn and
Sham of the Schr\"odinger kinetic energy operator is variationally
correct, while the equivalent local potential required for a valid
Thomas-Fermi theory, a Fr\'echet functional derivative of the Kohn-Sham
ground-state kinetic energy functional, does not exist.  The argument
of Holas and March is clearly invalid for the simple example of the
lowest triplet state of a two-electron atom with noninteracting 
electrons.  Why this fails, as do earlier arguments in the literature,
has been explained in recent publications, summarized here.
\end{abstract}
\vspace*{5mm}
\par
The principal issue in dispute is the question whether an exact
{\it ab initio} Thomas-Fermi theory (TFT) exists for ground states,
equivalent to  Kohn-Sham density-functional theory (DFT).
This depends on the definition of a ground-state kinetic energy
functional of the electron density and on the definition and
computability of its density-functional derivative.  Holas and March
claim that a local (Fr\'echet) functional derivative of the Kohn-Sham
ground-state kinetic energy functional exists, as a rigorous  
consequence of variational theory.  This cannot be true, because it 
would imply the equivalence of ground-state TFT and DFT, which
result in clearly inconsistent variational equations for systems
with more than two electrons\cite{NES98}.
\par The conclusions of the cited paper\cite{NES01} are misstated by
Holas and March\cite{HAM02}.  In contrast to their allusion to "...
Nesbet's worrisome claims...", which they hold to be contrary to the 
accepted opinion that "...DFT is fundamentally correct and internally
consistent...", the analysis summarized in\cite{NES01} concludes 
that Kohn-Sham theory correctly uses the kinetic energy operator
${\hat t}=-\half\nabla^2$ of Schr\"odinger, and is variationally
correct in the local-density approximation(LDA).
However, there is no formal proof that a local
(Fr\'echet) functional derivative exists in general for the 
exchange-correlation density functional, nor for the Kohn-Sham kinetic
energy functional.  In fact, the latter Fr\'echet derivative cannot
exist in a valid quantum theory for more than two electrons.  This
precludes an exact TFT for ground states.  Because Holas and 
March cite standard arguments which lead to similar inconsistencies,
it is important to trace the point of failure of these arguments. This 
will be done here.
\par 
The fundamental incompatibility of Thomas-Fermi theory (TFT) and
Kohn-Sham theory (DFT) for more than two electrons has not been widely
recognized.  If the argument of Holas and March and earlier conclusions
in accepted literature were sound, TFT would be correct. 
If the two theories are incompatible, then DFT must fail,
in sharp contrast to the history of applications of these theories.
To illustrate this incompatibility, consider the lowest $^3S$ state of
an atom with two noninteracting electrons.  Kohn-Sham theory solves the
pair of Schr\"odinger equations
\begin{eqnarray}\label{KSeq}
{\hat t}\phi_{1s}&=&\{\epsilon_{1s}-v({\bf r})\}\phi_{1s}
\nonumber\\
{\hat t}\phi_{2s}&=&\{\epsilon_{2s}-v({\bf r})\}\phi_{2s}.
\end{eqnarray}
The orbital wave functions are orthogonal because the energy eigenvalues
are unequal.  They are independently normalized to unity and are equally
occupied.  An equivalent TFT would use a single Lagrange multiplier 
$\mu$ for the global normalization constraint $\int\rho d^3{\bf r}=2$.
If a local density functional derivative of the kinetic energy 
functional $T$ exists in the form 
$\frac{\delta T}{\delta\rho({\bf r})}=v_T({\bf r})$,
this implies the Thomas-Fermi equation
\begin{eqnarray}\label{TFeq}
v_T({\bf r})=\mu-v({\bf r}),
\end{eqnarray}
Eqs.(\ref{KSeq}) cannot be deduced from Eq.(\ref{TFeq}), because
two independent Lagrange multipliers cannot be derived from one.
Unconstrained by the exclusion principle, this Thomas-Fermi equation has
a ground-state solution in which the partial density $\rho_{1s}$ is
normalized to $2$, while $\rho_{2s}$ is normalized to zero.  The
total energy, $2\epsilon_{1s}$, is below the correct ground state.
The theory fails because a single global Lagrange multiplier $\mu$ does
not suffice to normalize $\rho_{1s}$ and $\rho_{2s}$ separately.  It
cannot replace the set of independent Lagrange multipliers $\epsilon_i$ 
required for Fermi-Dirac electrons.  Any theory that produces a 
Fr\'echet derivative of a kinetic energy functional implies incorrect 
physics unless all $\epsilon_i$ are equal. 
\par 
Holas and March present an argument which, if correct, would support the
conclusion that solution of a Thomas-Fermi equation and of Kohn-Sham
equations for the same variational model must lead to the same results.
This is obviously false for the simple model cited above.  Eqs.(9) of 
Holas and March give the functional differential $\delta T$ of the
defining Schr\"odinger orbital functional $T=\sum_in_i(i|{\hat t}|i)$. 
Because $\rho=\sum_in_i\rho_i=\sum_in_i\phi^*_i\phi_i$ in DFT, a sum of 
orbital densities,
\begin{eqnarray*}
\delta T=
 \sum_in_i\int d^3{\bf r}\{\epsilon_i-v({\bf r})\}\delta\rho_i({\bf r})
\end{eqnarray*}
for variations about solutions of the noninteracting Schr\"odinger
equations.  This is valid for all density variations induced by orbital 
variations, whether or not the orbitals are normalized\cite{NES01}.
These equations demonstrate that partial variations $\delta_i T$ of the
orbital functional $T$ about stationary states are unique functionals
of the partial density variations $\delta\rho_i$\cite{NES01a}.  This 
implies that the Kohn-Sham ground-state functional $T_s[\rho]$ can be
extended to include such infinitesimal variations and defines a density 
functional derivative for each orbital subshell\cite{NES01,NES01a},
\begin{eqnarray*}
\frac{\delta T}{n_i\delta\rho_i({\bf r})}=\epsilon_i-v({\bf r}).
\end{eqnarray*}
This is a G\^ateaux functional derivative, the generalization to
functional analysis of a partial derivative\cite{BAB92}.  A Fr\'echet
derivative, equivalent to a local potential function $v_T$, exists only
if all $\epsilon_i$ are equal\cite{NES01a}.
\par The Kohn-Sham density functional $T_s$ is obtained by evaluating 
$T$ for ground-state orbital functions, restricting functional $T_s$ to 
normalized ground states of otherwise arbitrary external local potential
functions.  The fallacy in the argument of Holas and March is that they 
cannot deduce the functional derivative from this functional, limited to
normalized orbitals and densities.  They modify Eqs.(9) by replacing the
parameters $\epsilon_i$, implied by the Schr\"odinger equation, by the
single parameter $\mu$, appropriate to TFT.  This leads to their 
Eq.(11), which is not consistent with Eqs.(9) unless all integrals 
$\int\delta\rho_i$ vanish, as they do if all $\rho_i$ are separately 
normalized.  Since this requires independent
Lagrange multipliers $\epsilon_i$, replacing them by the single
parameter $\mu$ is not justified unless all $\epsilon_i$ are equal. 
For fixed orbital normalization, all integrals of
constants drop out, so that Eqs.(9) and (11) cannot be distinguished. 
No conclusion can be drawn about functional derivatives unless the 
definition of $T_s$ is extended to density variations driven by orbital
variations that are unconstrained in the orbital Hilbert space. 
\par
Lagrange multiplier formalism requires the relevant functional 
to be defined for variations unconstrained by 
normalization\cite{NES02a}.  Otherwise Euler-Lagrange variational 
equations cannot be derived.  This motivates the extension of $T_s$ as
defined by Kohn and Sham to infinitesimal function-neighborhoods of
ground-state densities $\rho$, generated by infinitesimal variations of
the orbital functions $\phi_i$\cite{NES01a}.  For noninteracting 
electrons, Fermi-Dirac statistics requires separate normalization of
each electronic subshell density, which implies independent Lagrange
multipliers $\epsilon_i$ indexed by orbital subshells.  Unless all 
these orbital energies are equal, they cannot be replaced by a single
parameter $\mu$.  TFT uses such a global parameter to normalize the 
total density, with no reference to shell structure.  This is the
reason for the historic failure of TFT to describe electronic shell
structure\cite{NES01}.
\par For stationary states, the G\^ateaux derivative implied by Eqs.(9)
is operationally equivalent to the linear operator ${\hat t}$, because
$(\epsilon_i-v)\phi_i={\hat t}\phi_i$\cite{NES01}.  It can be expressed
as a local mean orbital kinetic energy,
\begin{eqnarray*}
\frac{\delta T}{n_i\delta\rho_i({\bf r})}=
   \frac{\phi^*_i({\bf r}){\hat t}\phi_i({\bf r})}{\rho_i({\bf r})}.
\end{eqnarray*}
For independently normalized partial densities $\rho_i$, this determines
modified Thomas-Fermi equations equivalent to the usual Kohn-Sham or
Schr\"odinger equations for noninteracting electrons\cite{NES02}.
This restores Fermi-Dirac statistics and orbital shell structure to
Thomas-Fermi theory. 
\par Holas and March cite the work of Englisch and Englisch
\cite{EAE84a,EAE84b} as supporting their argument.  This is incorrect,
because the rigorous discussion of\cite{EAE84a,EAE84b} cannot 
distinguish between Fr\'echet and G\^ateaux functional derivatives.
Only normalized density functions are considered, and 
constants in the Euler-Lagrange equations are not determined.
Englisch and Englisch\cite{EAE84a} prove the existence in general of
a G\^ateaux derivative.  They do not consider the subshell structure
of the DFT density function, but arbitrarily insert a non-indexed single
parameter $\lambda$ into their Eq.(4.1), which if correct would imply
an exact TFT.  As shown above for the simple example of the 
lowest $1s2s^3S$ state of an atom with two noninteracting electrons,
this result is physically incorrect.  Construction of a G\^ateaux
derivative consistent with the operator ${\hat t}$ in noninteracting
Schr\"odinger equations and in Kohn-Sham equations uses only the
mathematics of variational theory in the orbital Hilbert space, 
following a long tradition of textbook derivations\cite{NES02a}.

\end{document}